\documentclass[a4paper,fleqn,usenatbib]{mnras}

\usepackage{newtxtext,newtxmath}

\usepackage{color}
\usepackage{amsmath}
\usepackage{graphicx}
\usepackage{amssymb}
\usepackage{psfrag}

\usepackage{etoolbox}
\makeatletter
\patchcmd\@combinedblfloats{\box\@outputbox}{\unvbox\@outputbox}{}{%
   \errmessage{\noexpand\@combinedblfloats could not be patched}%
}%
 \makeatother

\usepackage{color}
\usepackage{amsmath}
\usepackage{graphicx}
\usepackage{amssymb}
\usepackage{psfrag}

\newcommand{\beq}{\begin{equation}}
\newcommand{\eeq}{\end{equation}}
\newcommand{\nn}{\nonumber}
\newcommand{\rmd}{\mathrm{d}}
\newcommand{\brac}[1]{\left({#1}\right)}

\newcommand{\bB}{{\boldsymbol B}}
\newcommand{\bOm}{\boldsymbol\Omega}

\newcommand{\clF}{\mathcal{F}}
\newcommand{\Rly}{\mathfrak{R}_{LY}}

\title[Magnetar rotation and GWs]{Magnetar birth: rotation rates and gravitational-wave emission}
\author[S. K. Lander and D. I. Jones]{S. K. Lander${}^{1,2}$\thanks{samuel.lander@uea.ac.uk},
  D. I. Jones${}^3$\\ \\
         ${}^1$School of Physics, University of East Anglia, Norwich, NR4 7TJ, U.K.\\
         ${}^2$Nicolaus Copernicus Astronomical Centre, Polish Academy of Sciences, Bartycka 18, 00-716 Warsaw, Poland,\\
         ${}^3$Mathematical Sciences and STAG Research Centre, University of Southampton, Southampton SO17 1BJ, U.K.}

\begin{document}

\pagerange{\pageref{firstpage}--\pageref{lastpage}} \pubyear{0000}
\maketitle

\label{firstpage}

\begin{abstract}
Understanding the evolution of the angle $\chi$ between a magnetar's
rotation and magnetic axes sheds light on the star's birth
properties. This evolution is coupled with that of the
stellar rotation $\Omega$, and depends on the
competing effects of internal
viscous dissipation and external torques. We study this coupled
evolution for a model magnetar with a strong internal toroidal field, extending previous work by
modelling -- for the first time in this context -- the strong
proto-magnetar wind acting shortly after birth. We also account
for the effect of buoyancy forces on viscous dissipation at late times. Typically we find that
$\chi\to 90^\circ$ shortly after birth, then decreases towards
$0^\circ$ over hundreds of years. 
From observational indications that magnetars typically have small
$\chi$, we infer that these stars are subject to a stronger average exterior
torque than radio pulsars, and that they were born
spinning faster than $\sim 100-300$ Hz. Our results allow us to make
quantitative predictions for the gravitational and electromagnetic signals from a
newborn rotating magnetar. We also comment briefly on the possible connection with periodic
Fast Radio Burst sources.
\end{abstract}

\begin{keywords}
stars: evolution -- stars: interiors -- stars: magnetic fields -- stars: neutron -- stars: rotation
\end{keywords}

\maketitle

\section{Introduction}

Magnetars contain the strongest long-lived magnetic fields known in
the Universe. Unlike radio pulsars, the canonical neutron stars (NSs),
magnetars do not have enough rotational energy to power their
emission, and so the energy reservoir must be magnetic
\citep{TD95}. Through sustained recent effort in modelling, 
we now have a reasonable idea of the physics of the observed
\emph{mature} magnetars.

The early life of magnetars is far more poorly
understood, although models of various phenomena rely on them being
born rapidly rotating. Indeed, the very generation of magnetar-strength
fields is likely to involve one or more physical mechanisms that
operate at high rotation frequencies $f$: a convective dynamo
\citep{TD93} and/or the magneto-rotational
instability \citep{rembiasz}. Uncertainties about how these effects operate at the
ultra-high electrical conductivity of proto-NS matter -- where the
crucial effect of magnetic reconnection is stymied -- could be
partially resolved with constraints on the birth $f$ of magnetars.
In addition, a rapidly-rotating newborn magnetar could be the central
engine powering extreme electromagnetic (EM) phenomena -- 
superluminous supernovae and gamma-ray bursts (GRBs)
\citep{thomp04,kasen_bild,woosley10,M11}. Such a source might also emit
detectable gravitational waves (GWs)
\citep{cutler02,stella,dallosso09,kashiyama_etal_16}, though
signal-analysis
difficulties \citep{dallosso18} make it particularly
important to have realistic templates of the evolving star. As we will
see later, detection of such a signal would provide valuable constraints
on the star's viscosity (i.e. microphysics) and internal magnetic field.

A major weakness in all these models is the lack of convincing observational
evidence for newborn magnetars with such fabulously high rotation
rates; the galactic magnetars we observe have spun down to rotational periods
$P\sim 2-12$ s \citep{ol-kaspi}, and heavy proto-magnetars formed through
binary inspiral may since have collapsed into black holes. Details of
magnetar birth are, therefore, of major
importance. In this paper we show that an evolutionary model of magnetar
inclination angles -- including, for the first time, the key effect of a neutrino-driven 
proto-magnetar wind -- allows one to infer details about their birth rotation,
GW emission, and the prospects for accompanying EM signals.
Furthermore, two potentially periodic Fast Radio Burst (FRB)
  sources have very recently been discovered
  \citep{aetal_20,retal_20}, which may be powered by young precessing
  magnetars \citep{lbb_20,zl_20}; we show that our work allows constraints to be
  put on such models.

%%%%%%%%%%%%%%%%%%%%%%%%%%%%%%%%%%%%%%%%%%%%%%%%%%%%%%%%%%%%
\section{Magnetar evolution}

We begin by outlining the evolutionary phases of interest here. We consider a magnetar a few seconds after birth, once
processes related to the generation and rearrangement of 
magnetic flux have probably saturated. The physics of each phase will
be detailed later.

\emph{Early phase ($\sim$ seconds)}: the proto-NS is hot and still partially
neutrino-opaque. A strong particle wind through the
evolving magnetosphere removes angular momentum from the star. Bulk
viscosity -- the dominant process driving internal dissipation -- is suppressed.

\emph{Intermediate phase ($\sim$ minutes--hours)}: now transparent
to neutrinos, the star cools rapidly,
and bulk viscosity turns on. The wind is now
ultrarelativistic, and the magnetospheric structure has settled.

\emph{Late phase ($\sim$ days and longer)}: the presence of buoyancy
forces affects the nature of fluid motions within the star, so that
they are no longer susceptible to dissipation via bulk viscosity. The star slowly cools and
spins down.

%%%%%%%%%%%%%%%%%%%%%%%%%%%%%%%%%%%%%%%%%%%%%%%%%%%%%%%%%%%%
\subsection{Precession of the newborn, fluid magnetar}

Straight after birth, a
magnetar (sketched in Fig. \ref{field_geom}) is a fluid body; its
crust only freezes later, as the star cools. Normally, the only steady
motion that such a fluid body can sustain is rigid rotation about one axis $\bOm$. However, the star's
internal magnetic field\footnote{Later on we will use $B_{\rm int}$ more
precisely, to mean the volume-averaged internal magnetic-field strength.} $B_{\rm int}$ provides a certain `rigidity'
to the fluid, manifested in
the fact that it can induce some distortion $\epsilon_B$ to the star
\citep{chand-fermi}. For a dominantly poloidal $B_{\rm int}$ this
distortion is oblate; whereas a dominantly
toroidal $B_{\rm int}$ induces a prolate distortion. If the magnetic axis $\bB$
is aligned with $\bOm$, the magnetic and centrifugal distortions will also be aligned, and the
stellar structure axisymmetric and stationary -- but if they are
misaligned by some angle $\chi$, the
primary rotation about $\bOm$ will no longer
conserve angular momentum; a slow secondary rotation with period
\beq
P_{\rm prec}=\frac{2\pi}{\Omega\epsilon_B\cos\chi}
\eeq
about $\bB$ is also needed. These two rotations
together constitute rigid-body free precession, but since the star is fluid this bulk precession must be
supported by internal motions \citep{spitzer,mestel72}.
The first self-consistent solution for these motions, requiring
second-order perturbation theory, was only recently completed \citep{LJ17}.

On secular timescales these internal motions undergo viscous
damping, and the star is subject to an external EM
torque \citep{mestel72,jones76}. The latter effect tends to drive
$\chi\to 0^\circ$, as recently explored by \citet{sasmaz} in the
context of newborn magnetars; and if the star's magnetic distortion is
oblate, viscous damping of the internal motions supporting precession
also causes $\chi$ to decrease. Viscous damping of a
prolate star (i.e. one with a dominantly toroidal $B_{\rm int}$) is
more interesting: it drives $\chi\to 90^\circ$, and thus competes
with the aligning effect of the exterior torque. Therefore, whilst it
is not obvious how the internal motions could themselves be directly visible,
the effect of their dissipation may be.

In our previous paper,
\citet{LJ18}, we presented the first study of the evolution of $\chi$
including the competing effects of the exterior torque and internal dissipation.
The balance between these effects was shown to be delicate -- and so it is
important to capture the complex physics of the newborn magnetar as faithfully as
possible. In attempting to do so, our calculation will resort to a number of
approximations and parameter-space exploration of uncertain
quantities. Nonetheless, as we will discuss at the end, we believe our
conclusions are generally insensitive to this uncertainties -- and
that confronting these issues is better than ignoring them.

%%%%%%%%%%%%%%%%%%%%%%%%%%%%%%%%%%%%%%%%%%%%%%%%%%%%%%%%%%%%
\subsection{The evolving magnetar magnetosphere}
\label{wind_deriv}

\begin{figure}
\begin{minipage}[c]{\linewidth}
\begin{center}
\includegraphics[width=0.8\textwidth]{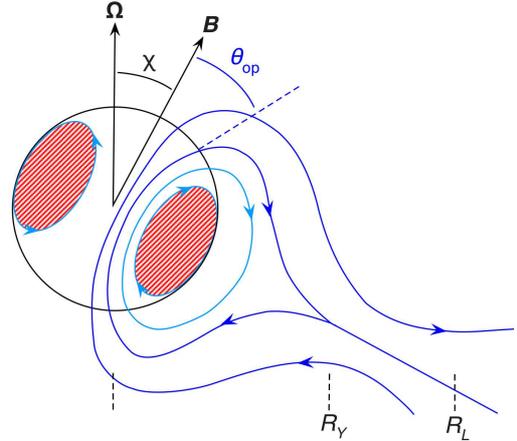}
\caption{\label{field_geom} Interior and exterior field of a newborn
  magnetar. Poloidal-field lines are shown in blue; the internal
  toroidal field (directed perpendicular to the page) is located in the
  red shaded region. The exterior field geometry and the star's
  spindown depend on the
rotation and magnetic-field strength. The open-field line region of
the magnetosphere, with opening half-angle $\theta_{\rm op}$, begins at a line joining to an
equatorial current sheet at the $Y$-point, located at a
radius $R_Y$ from $\bOm$. Both $R_Y$ and the Alfv\'en radius $R_A$
evolve in time towards the light-cylinder radius $R_L$, with $R_Y\leq
R_A\leq R_L$.}
\end{center}
\end{minipage}
\end{figure}

The environment around a NS determines how rapidly it loses angular
momentum, and hence spins down. This occurs even if the exterior
region is vacuum, through Poynting-flux losses at a rate (proportional
to $\sin^2\chi$) which may be solved analytically
\citep{deutsch}. The vacuum-exterior assumption is still fairly frequently employed in the
pulsar observational literature, although it exhibits
the pathological behaviour that spin-down decreases as $\chi\to 0^\circ$ and
ceases altogether for an aligned rotator ($\chi=0^\circ$).

The magnetic-field structure outside a NS,
and the associated angular-momentum losses, change when one
accounts for the distribution of charged particles that will naturally
come to populate the exterior of a pulsar \citep{gold_jul}. Solving for the
magnetospheric structure is now analytically intractable,
but numerical force-free solutions for the cases of $\chi=0^\circ$ \citep{CKF} and
$\chi\neq 0^\circ$ \citep{spit06} demonstrate a structure similar to that
sketched in Fig. \ref{field_geom}: one region of closed, corotating equatorial
field lines and another region of `open' field lines around the polar
cap. The two are delineated by a separatrix: a cusped field line that joins an equatorial
current sheet at the $Y$-point $R_Y$. Corotation of particles along
magnetic fields ceases to be possible if their linear velocity exceeds
the speed of light; this sets the light cylinder radius
$R_L=c/\Omega$. In practice, simulations employing force-free electrodynamics find
magnetospheric structures with $R_Y=R_L$, although solutions with $R_Y<R_L$ are not, a priori,
inadmissible. The angular-momentum losses from these models proved to
be non-zero in the case $\chi=0^\circ$, in contrast with the vacuum-exterior case.
These losses again correspond to the radiation of Poynting flux, but
are enhanced compared with the vacuum case, since there is now
additional work done on the charge distribution outside the star
\citep{timok}. Results from these simulations should be applicable in the
ultrarelativistic wind limit, and since it appears $R_Y=R_L$
generically for this
case, the losses are also independent of any details of the magnetospheric structure.

Shortly after birth, however, a magnetar exterior is unlikely to
bear close resemblance to the standard pulsar-magnetosphere models.
A strong neutrino-heated wind of charged particles will carry angular
momentum away from the star \citep{thomp04} -- a concept familiar from the study of
non-degenerate stars \citep{schatz} -- and these
losses may dominate over those of Poynting-flux type. At large
distances from the star, a particle carries away more angular momentum
than if it were decoupled from the star at the stellar surface. At
sufficient distance, however, there will be no additional enhancement to angular momentum
losses as the particle moves further out; the wind speed exceeds the
Alfv\'en speed, meaning the particle cannot be kept in corotation with
the star. The radius at which the two speeds become equal is the
Alfv\'en radius $R_A$.

An additional physical mechanism for angular-momentum loss becomes
important at rapid rotation: as well as thermal pressure, a
centrifugal force term assists in driving the particle wind. Each
escaping particle then carries away an enhanced amount of angular momentum
\citep{mestel68a,mestel-spr}. The mechanism is active up to the sonic radius
$R_s=(GM/\Omega^2)^{1/3}$, at which these centrifugal forces
are strong enough to eject the particle from its orbit. If it is
still in corotation with the star until the point when it is
centrifugally ejected, i.e. $R_A\geq R_s$, the maximal amount of
angular momentum is lost.

Another source of angular-momentum losses is plausible in the
aftermath of the supernova creating the magnetar: a magnetic
torque from the interaction of the stellar magnetosphere with fallback
material. The physics of this should resemble that of the classic
problem of a magnetic star with an accretion disc \citep{ghosh_lamb},
but the dynamical aftermath of the supernova is far messier,
and results will be highly sensitive to the exact physical conditions
of the system. Attempting to account for fallback matter would
therefore not make our model any more quantitatively accurate.

We recall that there are four radii of importance in the magnetar-wind
problem. Two of them,
$R_L$ and $R_s$, depend only on the stellar rotation rate. The others
are $R_Y$, associated with electromagnetic losses, and $R_A$,
associated with particle losses. We will need to account for how these
quantities, which both grow until reaching $R_L$, evolve over the
early phase of the magnetar's life. Finally, we also need to know, at
a given instant, the dominant physics governing the star's angular-momentum
loss. This is captured in
the wind magnetisation $\sigma_0$, the ratio of Poynting-flux to particle kinetic energy losses:
\beq\label{sigma}
\sigma_0=\frac{B_{\rm ext}^2\clF_{\rm op}^2 R_*^4\Omega^2}{\dot{M}c^3},
\eeq
where $\clF_{\rm op}$ is the fraction of field lines which remain open beyond $R_Y$
(see Fig. \ref{field_geom}) and $B_{\rm ext}$ is the surface field strength. Note that the limits
$\sigma_0\ll 1$ ($\sigma_0\gg 1$) correspond to non(ultra)-relativistic
winds.

At present there are neither
analytic nor numerical solutions providing a full description of the
proto-magnetar wind. In the absence of these, we will
adapt the model of \citet{M11} (hereafter M11), which at least attempts to
incorporate, semi-quantitatively,
the main ingredients that such a full wind solution should have. Based on their work, we have devised a simplified
semi-analytic model for the magnetar wind, capturing the same
fundamental wind physics but more readily usable for our
simulations. Our description of the details is brief, but
self-contained if earlier results are taken on trust; we denote
some equation X taken from M11 by (M11;X).

To avoid cluttering what follows with mass and radius
factors, we report equations and results for our fiducial magnetar
model with $R_*=12$ km and a mass $1.4 M_\odot$. We have, however,
performed simulations with a $15$-km radius, $2.4 M_\odot$ model, as a crude approximation to a
massive magnetar formed through binary inspiral \citep{giaco-perna},
finding similar results.

We start from the established mass-loss rate
$\dot{M}_\nu$ \citep{qian_woos} of a
non-rotating, unmagnetised proto-NS:
\beq\label{Mdot}
\dot{M}_\nu=-6.8\times 10^{-5}M_\odot\textrm{s}^{-1}
                       \! \brac{\! \frac{L_\nu}{10^{52}\textrm{erg s}^{-1}}\! }^{\!  5/3}
                       \!\brac{\! \frac{E_\nu}{10\textrm{ MeV}}\! }^{\!  10/3},
\eeq
where $M_\odot$ is the solar mass and $L_\nu$ and $E_\nu$ are the neutrino luminosity and energy per neutrino,
respectively. The idea will be to adjust this result to account for the
effects of rotation and a magnetic field.
From the simulations of \citet{pons99} (see M11 Fig. A1), we make the
following fits to the evolution of $L_\nu$ and $E_\nu$: 
\begin{align}
  \frac{L_\nu}{10^{52}\textrm{ erg s}^{-1}}
           &\approx 0.7\exp\brac{-\frac{t\textrm{ [s]}}{1.5}}+0.3\brac{1-\frac{t\textrm{ [s]}}{50}}^4\ ,\nn\\
  \frac{E_\nu}{10\textrm{ MeV}}
           &\approx 0.3\exp\brac{-\frac{t\textrm{ [s]}}{4}}+1-\frac{t\textrm{ [s]}}{60}\ .
\end{align}
Our model does not allow for evolution of the radius $R_*$, so our time zero corresponds to two seconds after
bounce, at which point $R_*$ has stabilised at $\sim 12$ km.

Charged particles can only escape the magnetised star along the
fraction of open field lines, so
the original mass-loss rate \eqref{Mdot} should be reduced to
$\dot{M}=\dot{M}_\nu \clF_{\rm op}$, where (M11;A4)
\beq
\clF_{\rm op}=1-\cos(\theta_{\rm op})
= 1- \cos\left[\arcsin\brac{\sqrt{R_*/R_Y}}\right].
\eeq
Now since $\cos(\arcsin x)=\sqrt{1-x^2}$, we have
\beq\label{f_op}
\clF_{\rm op}= 1-\sqrt{1-R_*\Rly/R_L},
\eeq
where $\Rly\equiv R_L/R_Y$. 
When $f\gtrsim 500$ Hz, the mass loss
may experience a centrifugal enhancement $\clF_{\rm cent}>1$, so that (M11;A15):
\beq\label{full_Mdot}
\dot{M}=\dot{M}_\nu \clF_{\rm op}\clF_{\rm cent}.
\eeq
Our approach will be first to ignore this
to obtain a slow-rotation solution, which we then use to
calculate $\clF_{\rm cent}$ (and hence the general $\dot{M}$) `perturbatively'.
We start by combining equations \eqref{f_op} and
\eqref{sigma} (with $\clF_{\rm cent}=1$) to get a relation between $\Rly$ and $\sigma_0$. But
another, phenomenological relation
$\mathfrak{R}_{LY}=\max\{(0.3\sigma_0^{0.15})^{-1},1\}$
\citep{bucc06,metz07} also links the two. The relations may therefore
be combined to eliminate $\sigma_0$:
\beq\label{Rly_eqn}
\brac{1-\sqrt{1-\frac{R_*}{R_L}\Rly}}\Rly^{1/0.15}
= \frac{0.3^{-1/0.15}c^3\dot{M}_\nu}{B_{\rm ext}^2 R_*^4\Omega^2}.
\eeq
This equation may be solved to find $\mathfrak{R}_{LY}$ for given
$B_{\rm ext},\Omega$ and $t$. It has real solutions as long as $R_*/R_Y<1$;
the $Y$-point cannot be within the star. As $R_Y\to R_*$ all
magnetospheric field lines become open, and the following limits are attained:
\beq
\mathfrak{R}_{LY}=R_L/R_*\ ,\ \clF_{\rm op}=1\ ,
\ \sigma_0=B_{\rm ext}^2 R_*^4\Omega^2/(\dot{M}_\nu c^3).
\eeq
Accordingly, in cases where equation \eqref{Rly_eqn} has no real solutions, we
use the above limiting values.

\begin{figure*}
\begin{center}
\begin{minipage}[c]{\textwidth}
\includegraphics[width=\textwidth]{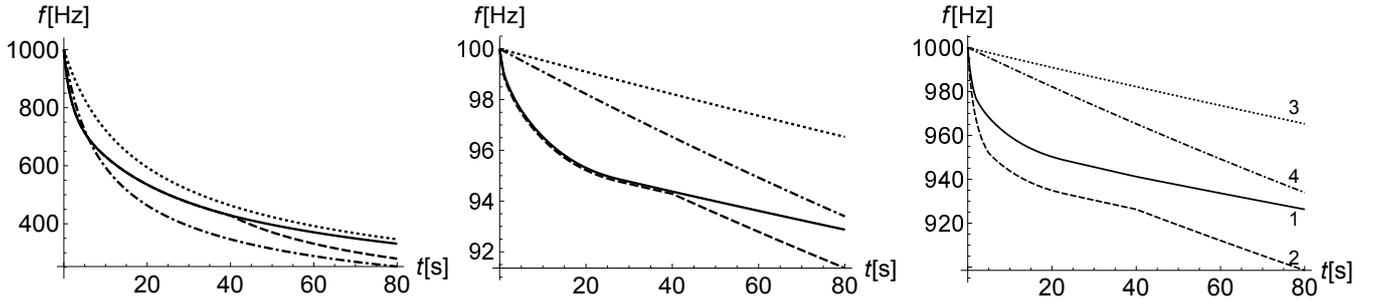}
\caption{\label{wind_nowind} The first 80 s of rotational evolution for different model
    newborn magnetars, with fixed $\chi$ and (from left to right):
    $f_0=1$ kHz, $B_{\rm ext}=10^{16}$ G; $f_0=100$ Hz, $B_{\rm
      ext}=10^{16}$ G; $f_0=1$ kHz, $B_{\rm ext}=10^{15}$ G.
    Linestyles 1,2 show (respectively) $\chi=0,90^\circ$ models evolved with the magnetar wind
    prescription described in section \ref{wind_deriv} for 40 s and
    thereafter with a `pulsar' prescription (third
    line of equation \eqref{Edot_wind}); linestyles 3,4 show
    the corresponding results using the `pulsar' prescription from birth.
    Note that before 40 s, lines 1 and 2 become indistinguishable from one another
    for higher $f_0$ and $B_{\rm ext}$.}
\end{minipage}
\end{center}
\end{figure*}

Next we move on to calculate the centrifugal enhancement. As discussed
earlier, this depends strongly on the location of $R_A$ with respect
to $R_s$. Only the former quantity depends on the magnetospheric
physics, and as for the $Y$-point location we find it convenient to work with the dimensionless radius
$\mathfrak{R}_{LA}\equiv R_L/R_A$. Now, M11 employ the
phenomenological relation $\mathfrak{R}_{LA}=\max\{\sigma_0^{-1/3},1\}$; we therefore just need
to find $\sigma_0$. To do so, we use the solution we have just
obtained for $\Rly$, plugging it in equation \eqref{full_Mdot} to make a
first calculation of $\dot{M}$ in the absence of any centrifugal enhancement
(i.e. setting $\clF_{\rm cent}=1$), then using the result in equation
\eqref{sigma} to find $\sigma_0$. We may now calculate the centrifugal
enhancement:
\beq
\clF_{\rm cent}=\clF_{\rm cent}^{\rm max}[1-\exp(-R_A/R_s)]+\exp(-R_A/R_s),
\eeq
where (M11;A12,A13)
\beq\label{fmax}
\clF_{\rm cent}^{\rm max}=\exp\left[\brac{\frac{f[\textrm{kHz}]}{2.8
\max\{\sin(\theta_{\rm op}),\sin\chi\}}}^{1.5}\right]
\eeq
is the maximum possible enhancement factor to the mass loss, occurring
when $R_A\geq R_s$.

The centrifugal enhancement relies on particles reaching large
distances from $\bOm$ whilst remaining in corotation; we can see this
will not happen if open field lines remain close to this axis out to large
distances. As a diagnostic of this, M11 assume that enhancement will
not occur if a typical open-field line angle 
$(\chi+\theta_{\rm op})\ll \pi/2$, but will do if $(\chi+\theta_{\rm
  op})\gtrsim\pi/2$. In practice we have to decide on an angle
delineating the two regimes: we take $\pi/4$. Accordingly, we will adopt equation
\eqref{full_Mdot} for the full mass-loss rate, but set $\clF_{\rm cent}=1$ when $\chi+\theta_{\rm op}<\pi/4$.
We now re-calculate equation \eqref{sigma} to find the full
$\sigma_0$, and so the EM energy-loss rate (M11;A5):
\beq\label{Edot_wind}
\dot{E}_{\rm EM}=\begin{cases}
  c^2\dot{M} \sigma_0^{2/3} & \sigma_0<1\textrm{ and }t<40\textrm{s}\\
  \frac{2}{3}c^2\dot{M} \sigma_0 & \sigma_0\geq 1\textrm{ and }t<40\textrm{s}\\
  -\frac{R_*^2}{4c^3}\Omega^4 B_{\rm ext}^2(1+\sin^2\chi) & t\geq 40\textrm{s}.
\end{cases}
\eeq
Within one minute, the bulk of the star's neutrinos have escaped
and so the proto-magnetar wind weakens greatly. Here we take the wind to
be negligible after $40$ s, at which point we switch to a fit \citep{spit06} to
numerical simulations of pulsar magnetospheres,
corresponding to the ultrarelativistic limit of the wind (i.e. kinetic
losses being negligible). For all our models $\sigma_0$
becomes large and $R_Y\to R_L$ before the $40$-second mark at which we switch to this
regime; see M11 for more details.

Note that the first and second lines of equation \eqref{Edot_wind} are
formally correct only in the limits $\sigma_0\ll 1$ and $\sigma_0\gg
1$, respectively, with no such simple expressions existing for the
case $\sigma_0\sim 1$. Treating the latter case is beyond the
scope of the present work, so we simply switch between the first two
regimes of equation \eqref{Edot_wind} at $\sigma_0=1$. We do not expect this to introduce any serious
uncertainty in our work, however: the wind magnetisation makes a rapid
transition between the two limiting regimes over a timescale short
compared with the evolution of both $\chi$ and $\Omega$.

Fig. \ref{wind_nowind} shows sample evolutions, comparing the
  magnetar wind prescription with one often used for pulsars (and also
  used, with a slightly different numerical prefactor, in
  \citet{LJ18}). For the extreme case of $f_0=1$ kHz, $B_{\rm ext}=10^{16}$ G (left-hand panel), we see that
  the rotation rate has roughly halved after 40 s
  for all models --  although the most rapid losses are suffered by the model
  with $\chi=90^\circ$ and the \emph{pulsar} prescription. For less
  extreme cases (middle and right panels), however, the magnetar wind
  always gives the greatest losses. Finally, as expected from equation \eqref{Edot_wind}, we see
  that the value of $\chi$ often has less effect on the
  magnetar-wind losses than those from the pulsar prescription.

%%%%%%%%%%%%%%%%%%%%%%%%%%%%%%%%%%%%%%%%%%%%%%%%%%%%%%%%%%%%
\subsection{Buoyancy forces}

At a much later stage, another physical effect needs to be
modelled, related to the role of buoyancy forces on internal motions.

The proportions of different particles in a NS varies with
depth. If one moves an element of NS matter to a different
depth, chemical reactions act to re-equilibrate it with its
surroundings, on a timescale $\tau_{\rm chem}$. When the temperature
$T$ is high, $\tau_{\rm chem}\ll P_{\rm prec}$, so moving fluid
elements are kept in chemical equilibrium. Once the star has cooled sufficiently, however, reactions will have slowed down
enough for fluid elements to retain a different composition from their surroundings
\citep{LJ18}; they will therefore be subject to a buoyancy force due to the chemical
gradient \citep{RG92}. This force tends to suppress radial motion, and
hence will predominantly affect the compressible piece of the motions
\citep{mestel72,lasky-glam}. For this phase, one would ideally generalise the lengthy calculation
of \citet{LJ17} to include buoyancy forces, but this is very likely to
be intractable. In lieu of this, we will
simply impose that the motions become
divergence-free below some temperature $T_{\rm solen}$, which we
define to be the temperature for which
\beq
P_{\rm prec}=\tau_{\rm chem}=0.2\brac{\frac{T}{10^9\textrm{ K}}}^{-6}
\brac{\frac{\bar\rho}{\rho_{\rm nuc}}}^{2/3},
\eeq
taking the expression for $\tau_{\rm chem}$ from \citet{RG92}, and where
$\rho_{\rm nuc}$ is nuclear density and $\bar\rho$ the average
core density. $T_{\rm solen}$ is clearly
a function of $B_{\rm int}$ and $\Omega$; its typical value is
$10^9-10^{10}$ K. For $T<T_{\rm solen}$, bulk viscous dissipation
(depending on the compressibility of the internal motions) therefore
becomes redundant, and we shut it off in our evolutions, leaving only
the ineffective shear-viscous dissipation. Without significant
viscous damping, the star's proclivity towards
becoming an orthogonal rotator ($\chi=90^\circ$) is suppressed.

Our evolutionary model employs standard fluid physics, and cannot
therefore describe any effects related to the gradual formation of the
star's crust. The star's motion depends on 
distortions misaligned from the rotation axis; at late stages this may
include, or even be dominated by, elastic stresses in the crust. For the
magnetar-strength fields we consider, however, it is reasonable to
assume that magnetic distortions dominate. Our fluid
model of a magnetar's $\chi$-evolution should predict the correct
long-timescale trend, even if it cannot describe
short-timescale seismic features (see discussion).

Finally, as the star cools the core will form superfluid components,
and the interaction between these may provide a new coupling mechanism
between the rotation and magnetic-field evolution \citep{ruderman_98}. It is not clear
what effect -- if any -- this will have on the long-timescale
evolution of $\chi$.

%%%%%%%%%%%%%%%%%%%%%%%%%%%%%%%%%%%%%%%%%%%%%%%%%%%%%%%%%%%%
\section{Evolution equations}

We follow the coupled $\Omega-\chi$ evolution of a newborn magnetar with a
strong, large-scale toroidal $B_{\rm int}$ in its core -- the expected outcome of
the birth physics \citep{jones76,TD93}. For stability reasons \citep{tayler80} this
must be accompanied by a poloidal-field component, but we will assume
that within the star it is small enough to be ignored here (it also retains consistency
with the solution we have for the internal motions; \citet{LJ17}). We
assume there is no internal motion, and hence no dissipation, in the
outer envelope (the region that becomes the crust once the star has
cooled sufficiently).

One unrealistic feature of purely
toroidal fields is that $B_{\rm ext}=0$. As in \citet{LJ18}, we will
assume that the poloidal-field component -- negligible within the star
-- becomes significant as one moves further out, and links to a
substantial $B_{\rm ext}$ sharing the same symmetry axis as
  $B_{\rm int}$. We then express the magnetic ellipticity as:
\beq
\epsilon_B
 =-3\times 10^{-4}\brac{\frac{B_{\rm int}}{10^{16}\textrm{ G}}}^2
 =-3\times 10^{-4}\brac{\frac{B_{\rm int}}{B_{\rm ext}}}^2\brac{\frac{B_{\rm ext}}{10^{16}\textrm{ G}}}^2,
 \eeq
where the first equality comes from self-consistent solutions of the star's hydromagnetic
equilibrium \citep{LJ09} with a purely toroidal internal field, and
the second equality links this ellipticity to the exterior field
strength (somewhat arbitrarily) through the ratio $B_{\rm ext}/B_{\rm
  int}$. Note that the negative sign of $\epsilon_B$ indicates that
the distortion is prolate.

A typical model encountered in the literature (e.g. \citet{stella}) assumes a `buried'
magnetic field, with $B_{\rm ext}/B_{\rm int}\ll 1$,
although self-consistent equilibrium models with vacuum exteriors have
$B_{\rm ext}\sim B_{\rm int}$ \citep{LJ09}. The
results for $f$ and $\chi$ vary
little with the choice of this ratio, since it is mostly the exterior torque,
i.e. $B_{\rm ext}$, that dictates the last-phase evolution, and we therefore
set the ratio to unity for simplicity unless stated otherwise -- an upper limit for our model,
as $B_{\rm ext}/B_{\rm int}\gtrsim 1$ would be inconsistent with the
toroidal field dominating within the star. Only in section \ref{GW-EM} do
we explore varying this ratio, as the predicted gravitational and
electromagnetic emission \emph{are} affected by the relative strength
of the magnetic field inside and outside the star.

The $\Omega$-evolution is given by the simple, familiar expression:
\beq\label{Omegadot}
\dot\Omega=\frac{\dot{E}_{\rm EM}}{I\Omega},
\eeq
where $I$ is the moment of inertia, whilst the $\chi$-evolution involves an interplay
between viscous dissipation $\dot{E}_{\rm visc}$ of internal fluid motions, and external
torques:
\beq\label{chidot}
\dot\chi=\frac{\dot{E}_{\rm visc}}{I\epsilon_B\sin\chi\cos\chi\Omega^2}
                      +\frac{\dot{E}_{\rm EM}^{(\chi)}}{I\Omega^2}.
\eeq
Now, $\dot\chi$ should vanish for
$\chi=0^\circ,90^\circ$ \citep{mestel68b}. The $\dot{E}_{\rm EM}$ from
equation \eqref{Edot_wind} does not satisfy this, however; it
represents the spindown part of the full external torque, whereas
$\dot\chi$ depends on a torque component orthogonal to this.
As a simple fix that gives the
correct limiting behaviour of $\dot\chi$, we take
$\dot{E}_{\rm EM}^{(\chi)}=\sin\chi\cos\chi\dot{E}_{\rm EM}$ for $t<40$ s.
For the later phase, \citet{philippov} suggest the expression
\beq\label{E_EM_chi}
\dot{E}_{\rm EM}^{(\chi)}=\frac{R_*^2}{4c^3}\Omega^4 B_{\rm
  ext}^2 k\sin\chi\cos\chi,
\eeq
based on fits to numerical simulations, and finding $k\approx 1$ for
dipolar pulsar magnetospheres. This is a sensible result, since
setting $k=1$ in equation \eqref{E_EM_chi} gives the analytic result
for the case of a vacuum exterior.   Evolutions for a vacuum
  exterior were performed in \citet{LJ18}; we also considered
  pulsar-like models, but with an alignment torque that did not vanish
  as $\chi\to 0^\circ$. The present treatment improves upon this.

Although equation \eqref{E_EM_chi} reflects the physics of pulsar
magnetospheres, the coronae of magnetars have a different physical
origin and are likely to be complex multipolar structures,
which will in turn affect the alignment torque. Furthermore, there are hints that a magnetar corona
may lead to an enhanced torque, $k>1$,
compared with the pulsar case \citep{TLK,younes}. On the other hand, for relatively
modest magnetic fields ($B\sim 10^{14}$ G) these
coronae are likely to be transient features \citep{belo_thomp,L16};
whilst we may still think of $k$ as embodying the long-term average
torque, it therefore seems implausible for the appropriate value of $k$ to be
far larger than unity. In the absence of suitable quantitative results
for magnetars, here we will simply adopt equation \eqref{E_EM_chi} to
describe the alignment, but explore varying the torque prefactor $k$ to check how
strong the alignment torque needs to be for consistency with the
model.

Finally, the gravitational radiation reaction torque on the star -- like its electromagnetic
counterpart -- has an aligning effect on the $\bB$ and $\bOm$ axes. It
is given by a straightforward expression that could be included
in our evolutions; we neglect it, however, as one can easily show that
the GW energy losses \citep{cutler-jones} in \eqref{Omegadot} and
\eqref{chidot} are always negligible compared with $\dot{E}_{\rm EM}$
for the models we consider. For instance, for a star with
$B_{\rm ext}=10^{16}$ G and $f=1$ kHz, the ratio of GW-driven
spindown to EM Poynting-type spindown
is $\sim 10^{-4}$.  This ratio scales as $f^2 B_{\rm int}^4/B_{\rm ext}^2$, so would be even
smaller for more slowly spinning and less strongly magnetised
stars. Furthermore, we have not considered the torque enhancement
due to the magnetar wind, which would further reduce the ratio.

Viscosity coefficients have strong $T$-dependence, so this should also be accounted for. We
assume an isothermal stellar core (recall that we do not consider
  dissipation in the envelope/crust) with
\beq
\frac{T(t)}{10^{10}\textrm{ K}}=
\begin{cases}
 40-\frac{39}{40}t[\textrm{s}] & t\leq 40\textrm{ s},\\
 [1+0.06(t[\textrm{s}]-40)]^{-1/6} & t> 40\textrm{ s},
\end{cases}
\eeq
which mimics the differing cooling behaviour in the neutrino diffusion
and free-streaming regimes, with the latter expression coming from
\citet{pgw_06}. The isothermal assumption is indeed quite
reasonable for the latter case, though less so for the former (see, e.g.,
\citet{pons99}); the temperature may vary by a factor of a few in
the core at very early times.

In calculating the viscous energy losses $\dot{E}_{\rm visc}$ we assume the same well known forms for shear and bulk viscosity  as described in
\citet{LJ18}.  Whilst shear viscosity is always assumed to be
active (albeit inefficient), bulk viscosity is not.  We have already
discussed why we take it to be inactive at late times when $T<T_{\rm solen}$, but it
is also suppressed in the early era, whilst the proto-neutron star
matter is still partially neutrino-opaque and reactions are inhibited. Following \citet{lai01}, we
will switch on bulk viscosity once the temperature drops below
$3\times 10^{10}$ K.  Note that while we include the viscosity
mechanisms traditionally considered in such analyses as ours, other
mechanisms can act. Of possible relevance in the very early
  life of our star is the shear viscosity contributed by the neutrinos
  themselves (see e.g. \citet{gmj_15}).  We leave study of this
  to future work, merely noting for now that its inclusion would
  increase the tendency for our stars to orthogonalise.

Whatever its microphysical nature,  viscous dissipation acts on the star's internal fluid motions, for
which we use the only self-consistent solutions to date
\citep{LJ17}. We do not allow for any evolution of $B_{\rm int}$.

%%%%%%%%%%%%%%%%%%%%%%%%%%%%%%%%%%%%%%%%%%%%%%%%%%%%%%%%%%%%

\section{Simulations}

\begin{figure}
\begin{minipage}[c]{\linewidth}
\begin{center}
  \includegraphics[width=0.8\textwidth]{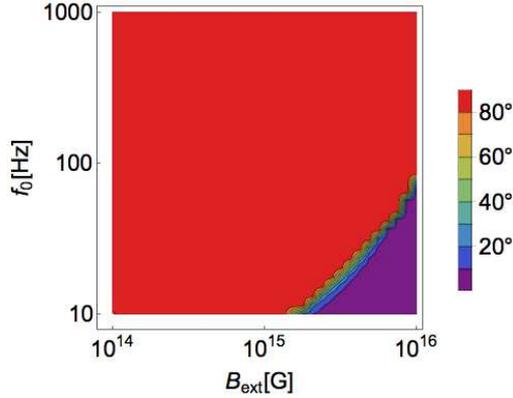}
\end{center}
\caption{\label{prerev} Distribution of inclination angles
  (colourscale) after one day, for a range of $f_0$ and $B_{\rm ext}$
  as shown, and with $\chi_0=1^\circ$ for all
  models. All models in the considered parameter range have already
  reached either the aligned- or orthogonal-rotator limit, though the
  orthogonal rotators will all start to align at later times.}
\end{minipage}
\end{figure}

We solve the coupled $\Omega-\chi$ equations
\eqref{Omegadot} and \eqref{chidot} with the physical input discussed
above.  The highly coupled and non-linear nature of the equations
means that numerical methods are required, and we therefore use adapted 
versions of the Mathematica notebooks described in detail in
\citet{LJ18}. Only in a few limits are
analytic results possible, e.g. at late times where $\chi$ has
reduced to nearly zero (see below), and the spin-down then proceeds
as the familiar power-law solution to equation (\ref{Edot_wind}).
Unless stated otherwise, we start all simulations with a
small initial inclination angle, $\chi_0\equiv\chi(t=0)=1^\circ$.

Fig. \ref{prerev} shows the distribution of $\chi$ after one day, for our chosen
newborn-magnetar parameter space
$f_0\equiv f(t=0)=10-10^3\textrm{ Hz},\ B_{\rm
  ext}=10^{14}-10^{16}\textrm{ G}$ and with $k=2$.
This is similar to our earlier results \citep{LJ18}, where the effect
of buoyancy forces on interior motions was not considered. As the
orthogonalising effect of internal viscosity becomes suppressed, the
orthogonal rotators can be expected to start aligning at later times, whilst
the small region of aligned rotators will obviously remain with
$\chi\approx 0^\circ$. If rapid rotation drives magnetic-field
amplification, however, a real magnetar born with such a low $f$ could
not reach $B\sim 10^{16}$ G.

\begin{figure}
\begin{minipage}[c]{\linewidth}
\begin{center}
  \includegraphics[width=0.85\textwidth]{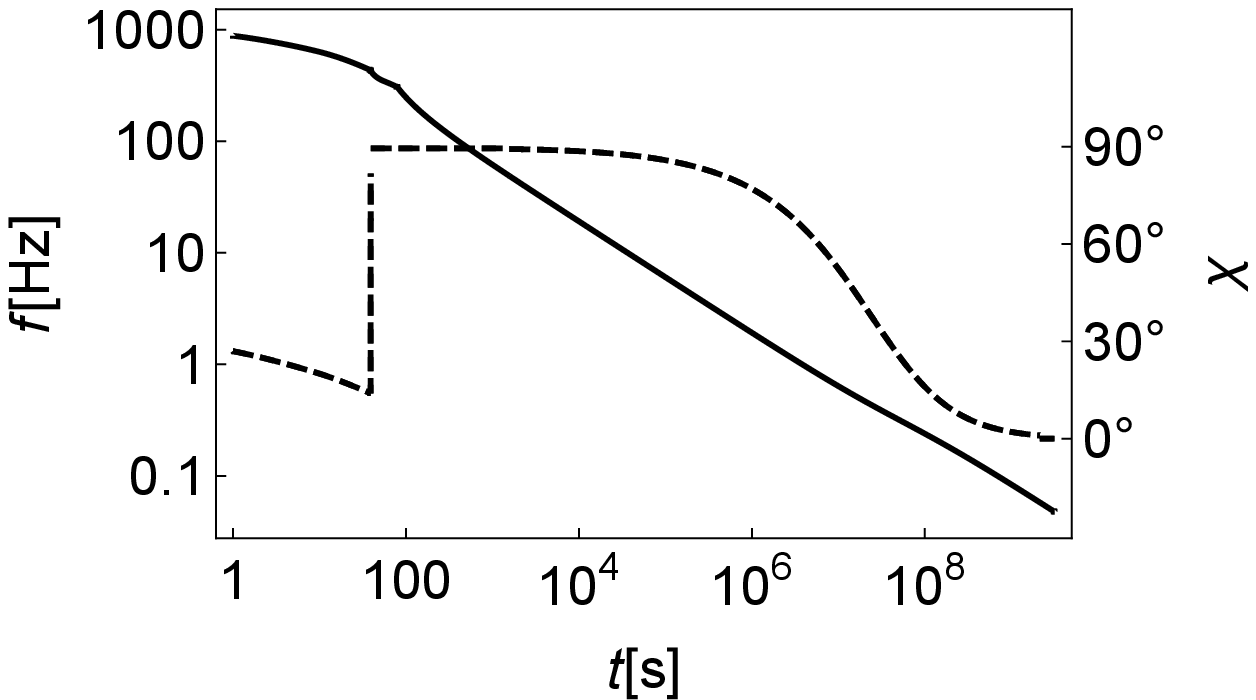}
  \includegraphics[width=0.85\textwidth]{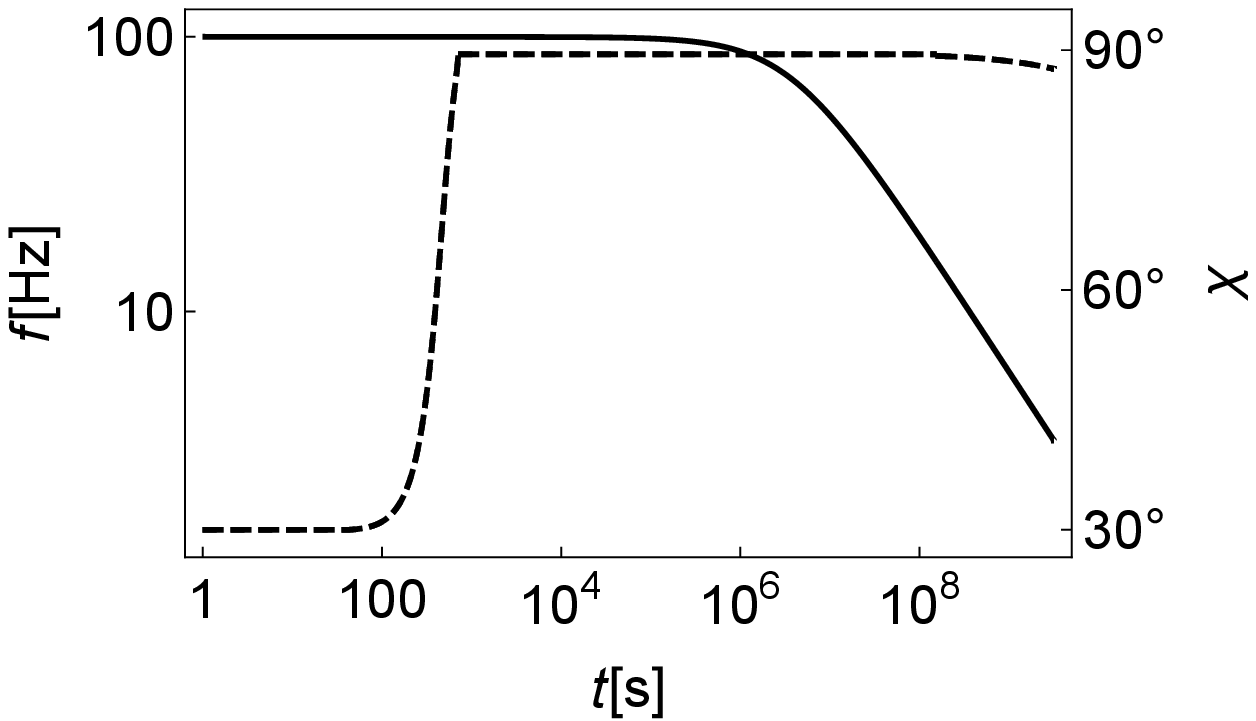}
\end{center}
\caption{\label{typical_evol} Evolution of $f$ (solid line) and $\chi$
  (dashed line) for two magnetars. Top: $f_0=1$ kHz,
  $B_{\rm int}=B_{\rm ext}=10^{16}$ G, bottom: $f_0=100$ Hz,
  $B_{\rm int}=B_{\rm ext}=10^{14}$ G. For illustrative purposes
  $\chi_0=30^\circ$ is chosen, but a smaller value is more likely. For
  both models $\chi$
  decreases for the first $\sim 40$ s, then increases rapidly to $90^\circ$
  as bulk viscosity becomes active, staying there until the internal
  motions become solenoidal (at $t\sim 10^3$ s for the left-hand
  model; at $t\sim 10^8$ s for the right-hand one), after which the spindown torque is able,
  slowly, to drive $\chi$ back towards $0^\circ$.}
\end{minipage}
\end{figure}

\begin{figure*}
\begin{center}
\begin{minipage}[c]{0.6\textwidth}
\includegraphics[width=\textwidth]{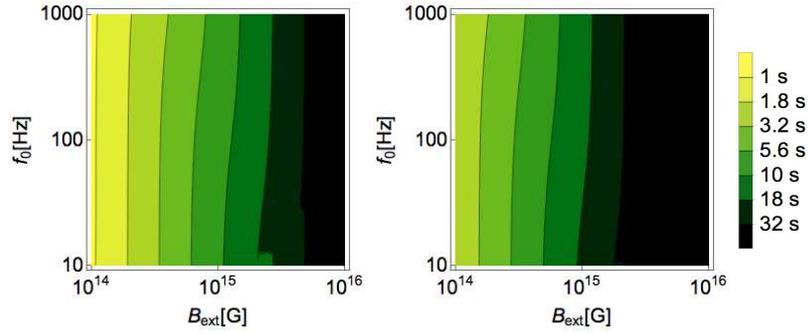}
\caption{\label{f_pres} Distribution of spin periods (colourscale) for
  magnetars with the shown range of $B_{\rm ext}$ and $f_0$ at ages
  of $1000$ yr (left) and $5000$ yr (right). For all models
  $\chi_0=1^\circ$ and $B_{\rm int}=B_{\rm ext}$.}
\end{minipage}
\end{center}
\end{figure*}

Fig. \ref{typical_evol} shows the way $f$ and $\chi$ evolve, for all
models in our parameter space except the aligned rotators of Fig. \ref{prerev}: an
early phase of axis alignment, rapid orthogonalisation, then slow
re-alignment. The evolution for most stars in our considered parameter
range is similar, though proceeds more slowly for lower $B_{\rm int}$, $B_{\rm
  ext}$ and $f_0$, as seen by comparing the left- and right-hand
panels (see also Fig. \ref{f_pres} and \ref{chi_pres}).

%%%%%%%%%%%%%%%%%%%%%%%%%%%%%%%%%%%%%%%%%%%%%%%%%%%%%%%%%%%%
\section{Comparison with observations}

Next we compare our model
predictions with the population of observed magnetars. Typical magnetars have $P\sim 2-12$ s and
$B_{\rm ext}\sim 10^{14}-10^{15}$ G; comparing these 
values with Fig. \ref{f_pres}, we see that they are consistent
with the expected ages of magnetars, roughly $1000-5000$ yr (see, e.g., \citet{tendulkar}). The
results in Fig. \ref{f_pres} are virtually insensitive to the exact
value of the alignment-torque prefactor $k$ (we take
$k=2$ in these plots). The model results are
very similar for different $B_{\rm int}$ and $\chi_0$, and the
vertical contours show that present-day periods are set primarily by
$B_{\rm ext}$, and give no indication of the birth rotation.

Observations contain more information than just $P$ and the
inferred $B_{\rm ext}$, however. The four magnetars observed in radio
\citep{ol-kaspi}: 
\begin{center}
\begin{tabular}{ccc}
    name                 & $P$/s & $B_{\rm ext}/(10^{14}\textrm{ G})$ \\
    \hline
  1E 1547.0-5408 & 2.1 & 3.2\\
  PSR J1622-4950 & 4.3 & 2.7 \\
  SGR J1745-2900 & 3.8 & 2.3 \\
  XTE J1810-197   & 5.5 & 2.1 \\
\end{tabular}\\
\end{center}
are particularly interesting. They have in common a flat spectrum and
highly-polarised radio emission that suggests they may all have a similar exterior geometry,
with $\chi\lesssim 30^\circ$ \citep{kramer07,camilo07,camilo08,levin12,shannon}. 
The probability of all four radio
magnetars having $\chi< 30^\circ$, assuming a random distribution of
magnetic axes relative to spin axes,
is $(1-\cos30^\circ)^4 \approx 3 \times 10^{-4}$, indicating that such a distribution is unlikely to happen by chance.
 Low values of $\chi$ could explain
the paucity of observed radio magnetars: if the emission is from the
polar-cap region, it would only be seen from a very favourable
viewing geometry. Beyond the four radio sources, modelling of magnetar
hard X-ray spectra also points to small $\chi$ \citep{belo13,hascoet},
giving further weight to the idea that small values of $\chi$ are
generic for magnetars.

\begin{figure*}
\begin{center}
\begin{minipage}[c]{0.8\linewidth}
\includegraphics[width=\textwidth]{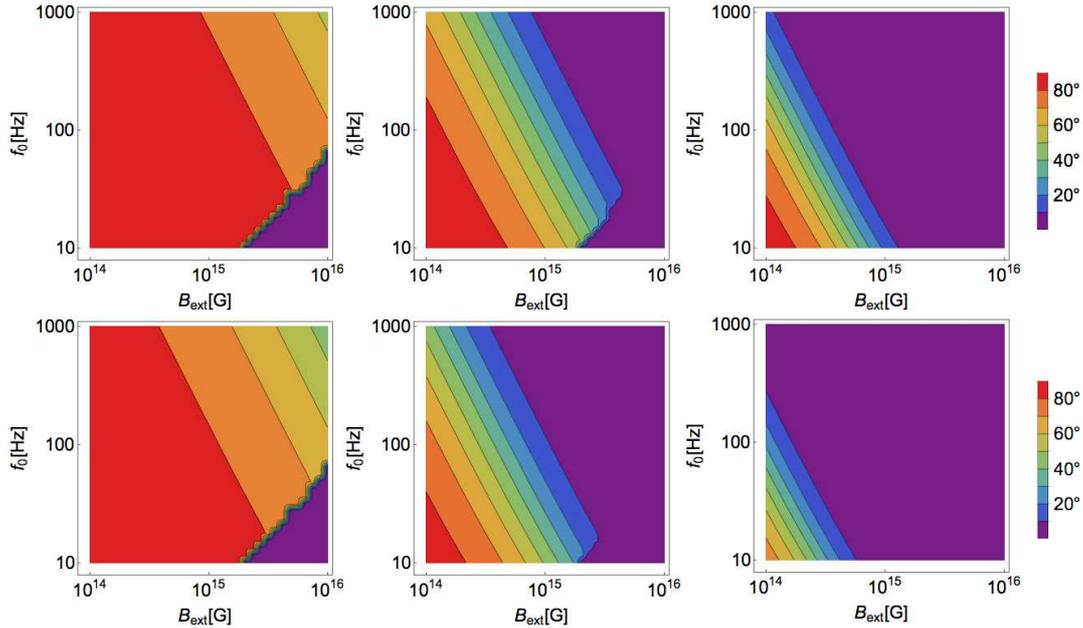}
\caption{\label{chi_pres}  Distribution of $\chi$ (colourscale) for magnetars with an alignment torque prefactor of
  (from left to right) $k=1,2$ and $3$; and at ages of $1000$ yr (top
  panels) and $5000$ yr (bottom panels). As before, $\chi_0=1^\circ$ and $B_{\rm int}=B_{\rm ext}$
  for all models.}
\end{minipage}
\end{center}
\end{figure*}

Now comparing with Fig. \ref{chi_pres}, we see that -- by contrast with the
present-day $P$ -- the present-day $\chi$ \emph{does} encode
interesting information about magnetar birth. Unfortunately, as noted
by \citet{philippov}, the results are quite sensitive to the
alignment-torque prefactor $k$. We are also hindered by the dearth of
reliable age estimates for magnetars. Nonetheless, we will still be able to draw
some quite firm conclusions, and along the way constrain the value of $k$.

Let us assume a fiducial mature magnetar
with $\chi<30^\circ$, $B_{\rm ext}=3\times 10^{14}$ G (i.e. roughly halfway between $10^{14}$ and
$10^{15}$ G on a logarithmic scale) and a strong internal toroidal field
(so that it will have had $\chi\approx 90^\circ$ at early times). We
first observe that such a star is completely inconsistent with $k=1$
unless it is far older than $5000$ yr, so we regard this as a
  strong lower limit.

If $k=2$, Fig. \ref{chi_pres} shows us that the birth rotation
must satisfy $f_0\gtrsim 1000$ Hz if our fiducial magnetar is $1000$
yr old, or $f_0\gtrsim 300$ Hz for a $5000$-yr-old magnetar. The former value
may just be possible, in that the break-up rotation rate is typically over 1
kHz for any reasonable neutron-star equation of state -- but is
clearly extremely high. The latter value of $f_0$ is more believable, but
  does require the star to be towards the upper end of the
  expected magnetar age range.

Finally, if $k=3$
the birth rotation is essentially unrestricted: it implies $f_0\gtrsim
20-100$ Hz for the age range $1000-5000$ yr. As discussed earlier,
however, this represents a very large enhancement to the torque --
with crustal motions continually regenerating the magnetar's corona --
and sustaining this over a magnetar lifetime
(especially $5000$ yr) therefore seems very improbable.

An accurate value of $k$ (or at least, its long-term average) cannot be determined without more
detailed work, so we have to rely on the qualitative arguments
above. From these, we tentatively suggest that existing magnetar observations indicate that
  $f_0\gtrsim 100-300$ Hz and $2\lesssim k< 3$ for these stars. Furthermore, from Fig. \ref{chi_pres},
we see that a single measurement of $\chi\gtrsim 15^\circ$ from one of
the more highly-magnetised (i.e. $B_{\rm ext}\sim 10^{15}$ G) observed magnetars would essentially rule
out $k\geq 3$.

%%%%%%%%%%%%%%%%%%%%%%%%%%%%%%%%%%%%%%%%%%%%%%%%%%%%%%%%%%%%

\section{Gravitational and electromagnetic radiation}
\label{GW-EM}

\subsection{GWs from newborn magnetars} \label{sect:GW}

\begin{figure}
\begin{minipage}[c]{\linewidth}
\begin{center}
  \includegraphics[width=0.9\textwidth]{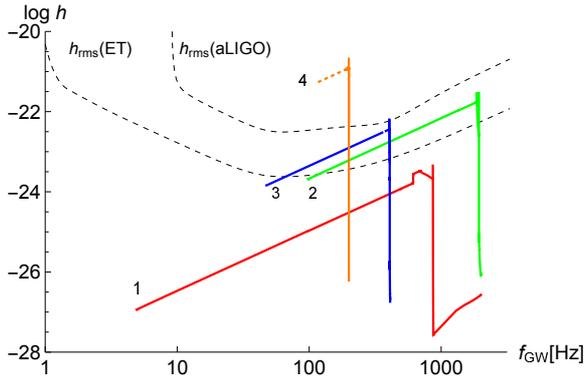}
\end{center}
\caption{\label{GW} GW signal $h_{\rm c}$ from four model newborn magnetars,
  against the noise curves $h_{\rm rms}$ for aLIGO and ET. Three models
  are for one week of signal at $d=20$ Mpc (i.e. Virgo galaxy cluster):
  (1) $f_0=1$ kHz, $B_{\rm ext}=B_{\rm int}=10^{16}$ G,
  (2) $f_0=1$ kHz, $B_{\rm ext}=0.05 B_{\rm int}=5\times 10^{14}$ G, 
  (3) $f_0=200$ Hz, $B_{\rm ext}=0.05 B_{\rm int}=10^{15}$ G;
  and the final signal (4) is at $d=10$ kpc (i.e. in our galaxy) and
  has duration of one year (solid line for the first week, dotted for
  the rest), with $B_{\rm ext}=0.05 B_{\rm int}=5\times 10^{13}$ G, $f_0=100$ Hz.}
\end{minipage}
\end{figure}

An evolution $\chi\to 90^\circ$ brings a NS into
an optimal geometry for GW emission \citep{cutler02}, and a few authors have
previously considered this scenario applied to newborn
magnetars \citep{stella,dallosso09}, albeit without the crucial
effects of the protomagnetar wind and self-consistent solutions for the
internal motions. By contrast, we have
these ingredients, and hence can calculate GWs from newborn magnetars
more quantitatively. In Fig. \ref{GW} we
plot the characteristic GW strain at distance $d$:
\beq
h_{\rm c}(t)=\frac{8G}{5c^4}\frac{\epsilon_B I\Omega(t)^2\sin^2\chi(t)}{d}\brac{\frac{f_{\rm GW}^2}{|\dot{f}_{\rm GW}|}}^{1/2}
\eeq
from four model magnetars with $\chi_0=1^\circ$, averaged over sky
location and source orientation, following \citet{jaran}. This signal is
emitted at frequency $f_{\rm GW}=2f=\Omega/\pi$. We also show the design rms
noise $h_{\rm rms}=\sqrt{f_{\rm GW} S_h(f_{\rm GW})}$ for the detectors aLIGO \citep{aLIGO_noise}
and ET-B \citep{ET_noise}, where $S_h$ is the detector's one-sided
power spectral density. Models 1 and 2 from Fig. \ref{GW} both have $f_0=1000$ Hz and
$B_{\rm int}=10^{16}$ G, but the former model has a much stronger
exterior field. As a result, it is subject to a strong wind torque,
which spins it down greatly before $\chi\to 90^\circ$, thus reducing
its GW signal compared with model 2.

Next we calculate the signal-to-noise ratio (SNR) for our selected
models, following \citet{jaran}:
\beq
\textrm{SNR}
=\left[\int\limits_{t=0}^{t_{\rm final}}\brac{\frac{h_{\rm c}}{h_{\rm
        rms}}}^2\frac{|\dot{f}_{\rm GW}|}{f_{\rm GW}}\ \rmd t\right]^{1/2}.
\eeq
Note that this expression assumes single coherent integrations.
In reality it will be difficult to track the evolving frequency well
enough to perform such integrations; see discussion in Section \ref{sect:discussion}.

Using aLIGO, models 1, 2, 3 and 4 have
$\textrm{SNR}=0.018,0.38,0.43,4.0$ for $t_{\rm final}=1$ week. With
ET, we find SNR values of
$0.19,4.5,4.4,47$ for models 1, 2, 3, 4, again taking
$t_{\rm final}=1$ week. Model 4 would be detectable for longer; taking
instead $t_{\rm final}=1$ 
yr gives $\textrm{SNR}=16$ ($200$) for aLIGO (ET).
Once $\chi$ for this model reduces below $90^\circ$, the GW signal
will gain a second harmonic at $f$, in addition to the one at
$2f$ \citep{jones_andersson_02} . However, even after $150$ yr (when the
model-4 signal drops below the ET noise curve), the star is still an
almost-orthogonal rotator, with $\chi=81^\circ$. In this paper,
therefore, it is enough to consider only the $2f$ harmonic.

Recently, \citet{dallosso18} studied GWs from newborn magnetars,
finding substantial SNR values even using aLIGO. To compare with
them, we take one of their $\textrm{SNR}=5$ models, which has
$B_{\rm ext}/B_{\rm int}=0.019$ and $f_0=830$ Hz. From their
equations (25) and (26), however, they appear to have a different
numerical prefactor from ours; if this was used in their calculations their
SNR values should be multiplied by $\sqrt{2/5}$ for direct comparison,
meaning the $\textrm{SNR}=5$ model would become $\textrm{SNR}\approx 3$. With our evolutions
we find $\textrm{SNR}\approx 2$ for the same model. This smaller value is to be
expected, since we account for two pieces of physics not present in
the \citet{dallosso18} model -- the magnetar wind and the aligning
effect of the EM torque -- which are both liable to reduce the GW signal.

%%%%%%%%%%%%%%%%%%%%%%%%%%%%%%%%%%%%%%%%%%%%%%%%%%%%%%%%%%%%
\subsection{Rotational-energy injection: jets and supernovae}

The rapid loss of rotational 
energy experienced by a newborn NS with very high $B_{\rm ext}$ and $f$ may be enough to power
superluminous supernovae, and/or GRBs. Because our wind model is
based on M11, our results for energy losses are similar to theirs, and the evolving $\chi$ only
introduces order-unity differences to the overall energy losses.  What
may change with $\chi$, however, is which phenomenon the lost rotational
energy powers: \citet{margalit} argue for a model with a partition of
the energy, predominantly powering a jet and GRB for
$\chi\approx 0^\circ$ and thermalised emission contributing to a more
luminous supernova for $\chi\approx 90^\circ$.

The amplification of a nascent NS's magnetic field to magnetar
  strengths is likely to require dynamo action, with differential
rotation playing a key role, and so we anticipate both poloidal and
toroidal components of the resulting
magnetic field to be approximately orientated around the rotation
axis. In this case, $\chi$ at birth would be
small -- and decreases further whilst the stellar matter is still
partially neutrino-opaque ($\sim 38$ s in our model). For all of
this phase we therefore find -- following
\citet{margalit} -- that most lost rotational energy manifests
itself as a GRB. Following this, the stellar matter becomes
neutrino-transparent and bulk viscosity activates, rapidly driving $\chi$
towards $90^\circ$. By this point $f$ will
have decreased considerably, but could still be well over $100$
Hz. The star remains with $\chi\approx 90^\circ$ for $\sim 10^6$ s in
the case of an extreme millisecond magnetar, or otherwise longer; see Fig. \ref{typical_evol}. Now the rotational
energy is converted almost entirely to thermal energy and ceases to
power the jet. Therefore, at any one point during the magnetar's
evolution, one of the two EM scenarios is strongly favoured.

\subsection{Fast Radio Bursts}

Finally, we will comment briefly on the periodicities that have
  been seen in two repeating FRB sources (to date).
  \citet{aetal_20} reported evidence for a $16$-day periodicity in FRB
  180916.J0158+65 over a data set of $\sim 1$ year, whilst \citet{retal_20}
  found somewhat weaker evidence for a $159$-day periodicity in FRB
  121102 from a $\sim 5$-year data set.  The possibility
  of magnetar precession providing the required periodicity was pointed out by
  \citet{aetal_20}, and developed further in \citet{lbb_20} and
  \citet{zl_20}, with the periodicity being identified with the free
  precession period.  

As noted by  \citet{zl_20}, the lack of a measurement of a spin period
introduces a significant degeneracy (between $P$ and $\epsilon_B$, in
our notation).  Nevertheless, a few common-sense considerations help
to further constrain the model.  In addition to reproducing the free
precession period, a successful model also has to predict no
significant evolution in spin frequency (as noted by \citet{zl_20}) or
in $\chi$, over the $\sim 1$--$5$  year durations of the observations.
Also, the precession
angle cannot be too close to zero or $\pi/2$, as otherwise there
would be no geometric modulation of the emission.
Finally, a requirement specific to the model of
  \citet{lbb_20} is that the magnetar should be only tens of years old.

Our simulations show that requiring $\chi$ to take an
  intermediate value is a significant constraint.  At sufficiently late
times the electromagnetic torque wins out, and the star aligns ($\chi
\rightarrow 0$), an effect not considered in either \citet{lbb_20}
or \citet{zl_20}. We clearly can accommodate stars of ages $\sim
10-100$ yr with such intermediate $\chi$ values; see the top panel of
Fig. \ref{typical_evol}. Such magnetars in this age range experience,
however, considerable spindown: from our evolutions we find a
decrease of around $4\%$ in the 
spin and precession frequencies over a year at age $10$ yr, and a
$0.5\%$ annual decrease at age $100$ yr. More work is clearly needed
to see whether this is compatible with the young-magnetar model, and
we intend to pursue this matter in a separate study.

%In the top panel, a highly magnetised star born fast is nearly aligned at
%age $\sim 30$ years, while a less strongly magnetised star born with a
%lower spin frequency has orthogonalised.  Clearly, there will exist a
%family of intermediate solutions, with intermediate $\chi$ values at
%ages $\sim 30$ years.  Identification  of this family, and imposition
%of the constraint of little evolution in spin and inclination angle on
%timescales of interest, is clearly required.  We intend to  pursue
%this matter, and will present out results in a separate study.  }

%%%%%%%%%%%%%%%%%%%%%%%%%%%%%%%%%%%%%%%%%%%%%%%%%%%%%%%%%%%%

\section{Discussion} \label{sect:discussion}

Inclination angles encode important information
about NSs that cannot be otherwise constrained. In particular, hints
that observed magnetars generically have small $\chi$ places a
significant and interesting constraint on their
rotation rates at birth, $f_0\gtrsim 100-300$ Hz, and shows that their
exterior torque must be stronger than that predicted for pulsar magnetospheres. More detailed modelling
of this magnetar torque may increase this minimum
$f_0$. Because our models place lower limits on $f_0$ (from the shape
of the contours of Fig. \ref{chi_pres}), they complement
other work indicating upper limits of $f_0\lesssim 200$ Hz, based on
estimates of the explosion energy from magnetar-associated supernovae
remnants \citep{vink_kuip}.

Typically, a newborn magnetar experiences an evolution where $\chi\to 90^\circ$
within one minute. At this point it emits its strongest GW signal.  
For rapidly-rotating magnetars born in the Virgo cluster, for which
the expected birth rate is $\gtrsim 1$ per year
\citep{stella}, there are some prospects for detection of this signal
with ET, provided that the ratio $B_{\rm ext}/B_{\rm int}$
is small. Such a detection would allow us to infer the unknown $B_{\rm
  int}$. A hallmark of the magnetar-birth
  scenario we study would be the onset of a signal with a delay of
  roughly one minute from the initial explosion. The delay is
  connected with the star becoming neutrino-transparent, and so
  measuring this might provide a probe of the newborn star's microphysics. Note, however, that the
actual detectability of GWs depends upon the signal analysis method employed -- most
importantly single-coherent verses multiple-incoherent integrations of
the signal -- and on the amount of prior information obtained from
EM observations, most importantly signal start time and
sky location. For a realistic search, reductions of sensitivity by a
factor of $5$-$6$ are possible \citep{dallosso18, miller_etal_18}.

Stronger magnetic fields do not necessarily improve
prospects for detecting GWs from newborn magnetars.  A strong $B_{\rm ext}$ causes a dramatic
initial drop in $f$ before orthogonalisation, resulting in a diminished GW signal. The lost
rotational energy from this phase will predominantly power a GRB, and
later energy losses may be seen through increased luminosity of the supernova. \emph{Less}
electromagnetically spectacular supernovae may therefore be better
targets for GW searches.

The birth of a NS in our galaxy\footnote{It is optimistic -- but not
  unreasonable -- to
  anticipate seeing such an event, with birth rates of maybe a few per
  century \citep{lorimer,faucher}.} need not have such extreme parameters
to produce interesting levels of GW emission, as long as it has a fairly strong internal
toroidal field, $B_{\rm int}\gtrsim 10^{14}$ G, and $f_0\gtrsim 100$
Hz. These are plausible birth parameters for a typical radio pulsar,
since $B_{\rm ext}$ will typically be somewhat weaker than $B_{\rm int}$. Such a star
will initially experience a similar evolution to that reported here,
but slower, giving the star time to cool and begin forming a
crust. Afterwards, the evolution of $\chi$ will probably proceed in a slow, stochastic way
dictated primarily by crustal-failure events: crustquakes or episodic
plastic flow. Regardless of the details of this evolutionary phase,
we find that the long-timescale trend for all NSs should be the alignment
of their rotation and magnetic axes, which is in accordance with
observations \citep{tauris_man,welt_john,john_kara_19}.

Many of our conclusions will not be valid for NSs whose magnetic fields are dominantly poloidal,
rather than toroidal. In this case the magnetically-induced distortion
is oblate, and there is no obvious mechanism for $\chi$ to increase;
it will simply decrease from birth. The expectation that all NSs
eventually tend towards $\chi\approx
0^\circ$ remains true, but our constraints on magnetar birth
would likely become far weaker and the GW emission from this phase
negligible. The lost rotational energy from the newborn magnetar would
power a long-duration GRB almost
exclusively, at the expense of any luminosity enhancement to the
supernova. Poloidal-dominated fields are,
however, problematic for other reasons: it is not clear how they would
be generated, whether they would be stable, or whether magnetar
activity could be powered in the absence of a toroidal field stronger
than the inferred exterior field. This aspect of the
life of newborn magnetars clearly deserves more detailed modelling.

%%%%%%%%%%%%%%%%%%%%%%%%%%%%%%%%%%%%%%%%%%%%%%%%%%%%%%%%%%%%
%%%%%%%%%%%%%%%%%%%%%%%%%%%%%%%%%%%%%%%%%%%%%%%%%%%%%%%%%%%%
%%%%%%%%%%%%%%%%%%%%%%%%%%%%%%%%%%%%%%%%%%%%%%%%%%%%%%%%%%%%

\section*{Acknowledgements}

We thank Simon Johnston and Patrick Weltevrede for valuable
discussions about inclination angles. We are also grateful to
Cristiano Palomba, Wynn Ho, and the referees
for their constructive criticism.  SKL acknowledges support from the European Union's Horizon 2020
research and innovation programme under the Marie Sk\l{}odowska-Curie
grant agreement No. 665778, via fellowship UMO-2016/21/P/ST9/03689 
of the National Science Centre, Poland. DIJ acknowledges support from
the STFC via grant numbers ST/M000931/1 and ST/R00045X/1.
Both authors thank the PHAROS COST Action (CA16214) for partial support.

\bibliographystyle{mnras}

%\newpage

\bibliography{references}

\label{lastpage}

\end{document}